\documentclass[preprint,aps,prd,superscriptaddress,nofootinbib,tightenlines]{revtex4}

\usepackage{amsfonts}
\usepackage{multirow}
\usepackage{mathrsfs}
\usepackage{graphicx}
\usepackage{amsmath}
\usepackage{amssymb}
\usepackage{bm}
\usepackage{bbm}
\usepackage{color}
\usepackage{array}
\usepackage{diagbox}
\usepackage{float}
\allowdisplaybreaks[4] 

\def\OMIT#1{}

\def\hlinew#1{%
  \noalign{\ifnum0=`}\fi\hrule \@height #1 \futurelet
   \reserved@a\@xhline}
\usepackage{array}
\newcommand{\PreserveBackslash}[1]{\let\temp=\\#1\let\\=\temp}
\newcolumntype{C}[1]{>{\PreserveBackslash\centering}p{#1}}
\newcolumntype{R}[1]{>{\PreserveBackslash\raggedleft}p{#1}}
\newcolumntype{L}[1]{>{\PreserveBackslash\raggedright}p{#1}}

\newcommand{\nn}{\nonumber}

\newcommand{\beq}{\begin{equation}}
\newcommand{\eeq}{\end{equation}}
\newcommand{\bqa}{\begin{eqnarray}}
\newcommand{\eqa}{\end{eqnarray}}

\newcommand\fverb{\setbox\fverbbox=\hbox\bgroup\verb}
\newcommand\fverbdo{\egroup\medskip\noindent%
			\fbox{\unhbox\fverbbox}\ }
\newcommand\fverbit{\egroup\item[\fbox{\unhbox\fverbbox}]}
\newbox\fverbbox

\makeatletter

\newcommand{\Rmnum}[1]{\expandafter\@slowromancap\romannumeral #1@}
\makeatother

\begin{document}
\title{\mbox{}\\[10pt]
$Z$-boson radiative decays to an $S$-wave quarkonium at NNLO and NLL accuracy}

\author{Wen-Long Sang~\footnote{wlsang@swu.edu.cn}}
 \affiliation{School of Physical Science and Technology, Southwest University, Chongqing 400700, China\vspace{0.2cm}}
\affiliation{College of Physics, Chongqing University, Chongqing 401331, China\vspace{0.2cm}}

\author{De-Shan Yang~\footnote{yangds@ucas.ac.cn}}
 \affiliation{School of Physical Sciences, University of Chinese Academy of Sciences, Beijing 100049, China\vspace{0.2cm}}

\author{Yu-Dong Zhang~\footnote{ydzhang@mails.ccnu.edu.cn}}
 \affiliation{Institute of Particle Physics and Key Laboratory of Quark and Lepton Physics (MOE),
 	Central China Normal University,Wuhan, Hubei 430079, China\vspace{0.2cm}}

\date{\today}

\begin{abstract}
Within the framework of nonrelativistic QCD (NRQCD) factorization formalism, we compute QCD next-to-next-to-leading order (NNLO) corrections to 
the helicity amplitudes as well as the decay width of $Z\to H+\gamma$, where $H$ can be $\eta_Q (Q=c,b), J/\psi$, or $\Upsilon$. In addition, we resum the next-to-leading logarithms (NLL) of ${m_Z^2}/{m_Q^2}$ to all orders of $\alpha_s$ for the leading-twist helicity amplitude by employing the light-cone factorization approach.  It is worth mentioning that we obtain the analytic expressions of the truncated NLL at $\alpha_s^2$.
We find that the $\mathcal{O}(\alpha_s)$ corrections are around 10\% for $\eta_c$ and $\Upsilon$ productions, however insignificant for $J/\psi$
and $\eta_b$ productions. The $\mathcal{O}(\alpha_s^2)$ corrections are moderate for charmonium production, while very small for bottomonium production. 
Moreover, it is found that the NLL resummation can considerably alter the NRQCD prediction, especially for $J/\psi$ production.
Combining the NRQCD and light-cone computation, we make phenomenological predictions on the decay widths and branching fractions.
In addition, we investigate the dependence of the theoretical results on the heavy quark mass, and find the branching fraction of
 $Z\to H+\gamma$ monotonically decreases as $m_Q$ increases.

\end{abstract}

\maketitle
\section{introduction}

It is an ideal platform to study the interplay between perturbative and nonperturbative QCD through the radiative decay of the $Z$ boson to a quarkonium. To date, experimentalists have made many attempts to search for such processes~\cite{ATLAS:2015vss,ATLAS:2015xkp,CMS:2018fzh}, yet have failed to find any signals. In recent years, several high-luminosity lepton colliders, such as {\tt ILC}~\cite{Baer:2013cma}, {\tt FCC-ee}~\cite{FCC:2018byv} and {\tt CEPC}~\cite{CEPCStudyGroup:2018ghi}, have been proposed to run at the $Z$ pole mass for a period of time. It will undoubtedly provide an opportunity to accumulate a large number of $Z$ bosons, thus increasing the chances to probe rare decay processes.

The exclusive processes $Z\to {\rm quarkonium}+\gamma$ have been extensively studied on the theoretical side, with the earliest computations dating back to the 1980s~\cite{Guberina:1980dc}. In Ref.~\cite{Luchinsky:2017jab}, Luchinsky studied these processes at lowest order in $\alpha_s$ and $v^2$ in both the nonrelativistic QCD (NRQCD)~\cite{Bodwin:1994jh} and light-cone (LC) factorization formalisms~\cite{Lepage:1980fj,Chernyak:1983ej}, where $v$ represents the typical velocity of the heavy quark in the quarkonium rest frame. In Ref.~\cite{Wang:2013ywc}, Wang {\it et al.} obtained the analytic expressions of the amplitudes for $Z\to {\rm quarkonium}+\gamma$ in the leading-power LC approximation at next-to-leading order (NLO) in $\alpha_s$.
Furthermore, Huang {\it et al.} presented calculations of the rates for $Z\to V+\gamma$ accurate up to the leading-power LC approximation at NLO both in $\alpha_s$ and $v$~\cite{Huang:2014cxa}, where $V$ denotes a vector quarkonium. 
Shortly afterwards, the resummation of the leading logarithms (LL) of $m_Z^2/m_Q^2$ for the decay width of $Z\to V+\gamma$ was carried out~\cite{Grossman:2015cak}. Bodwin {\it et al.} further considered the resummation of logarithms of $m_Z^2/m_Q^2$ for the $\mathcal{O}(\alpha_s)$ corrections as well as the $\mathcal{O}(v^2)$ corrections~\cite{Bodwin:2017pzj}. A combination of the next-to-leading logarithms (NLL) resummation and NLO fixed-order results was carried out for $\eta_Q+\gamma$ production in Ref.~\cite{Chung:2019ota}. The decay widths for $Z\to \Upsilon(nS)+\gamma$ have been calculated up to NLO in $\alpha_s$ based on the NRQCD, which are proposed to determine the $Zb\bar{b}$ coupling~\cite{Dong:2022ayy}. Very recently, the decay widths of $Z$ boson radiative decays to a $P$-wave quarkonium have been computed accurately up to next-to-next-to-leading order (NNLO) in $\alpha_s$ based on the NRQCD and LL resummation based on the LC approach~\cite{Sang:2022erv}. 
Moreover, the cross sections of $e^+e^-\to{\rm quarkonium}+\gamma$ at $Z$ factories have been computed in Refs.~\cite{Chen:2013mjb,Chen:2013itc,Sun:2014kva}. It is worth mentioning that some efforts toward the NNLO perturbative corrections to $e^+e^-\to {\rm quarkonium}+\gamma$ at B factories have been made in Refs.~\cite{Chen:2017pyi,Yu:2020tri,Sang:2020fql} in recent years.

In this work, we investigate the radiative decay of the $Z$ boson to a quarkonium $H$ ($H$ can be $\eta_Q$, $J/\psi$ or $\Upsilon$) 
by including both the NNLO perturbative corrections and the NLL resummation. We first 
compute the helicity amplitudes at NNLO in $\alpha_s$ and leading order (LO) in $v$ within the framework of NRQCD. To reduce the ambiguity in choosing the energy scale and uncertainty from the higher-order corrections arising from the large logarithms of $m_Z^2/m_Q^2$, we 
employ the LC formalism~\cite{Jia:2008ep} to refactorize the NRQCD short-distance coefficients (SDCs) and utilize the celebrated Efremov-Radyushkin-Brodsky-Lepage (ERBL) equation~\cite{Lepage:1980fj,Efremov:1979qk} to resum the large logarithms of $m_Z^2/m_Q^2$. Concretely, we will 
perform the NLL resummation for the leading-twist helicity amplitudes, i.e., resuming both the $\alpha_s^n\ln^n(m_Z^2/m_Q^2)$ and $\alpha_s^{n+1}\ln^n(m_Z^2/m_Q^2)$ to all orders of $\alpha_s$. 

The paper is organized as follows. In Sec.~\ref{sec-gen-for}, we present the theoretical framework to compute the 
decay widths of $Z\to H+\gamma$. In Sec.~\ref{sec-NRQCD}, we employ the NRQCD formalism to factorize the helicity amplitudes, 
introduce the procedure and techniques to compute the SDCs, and present the results of the helicity SDCs at various perturbative levels. Section~\ref{sec-LL} is devoted to the LC factorization for the leading-twist helicity SDCs. In addition, resummation of the NLL is formulated and explicitly carried out. The analytic expressions of the truncated  NLL at $\alpha_s^2$ are also obtained. A detailed phenomenological analysis is performed in Sec.~\ref{sec-phen}. Finally, we summarize in Sec.~\ref{sec-summary}. In Appendix~\ref{appendix-helicity-projectors}, we construct the helicity projectors. 
In Appendix~\ref{appendix-BL-kernels}, the explicit expressions for the Brodsky-Lepage (BL) kernels are given. 
In Appendix~\ref{appendix-convolutions}, we present some useful convolution formulas.

\section{theoretical framework for decay width ~\label{sec-gen-for}}

Applying the helicity amplitude formalism to analyze the hard exclusive production process proves to be convenient.
The unpolarized decay widths of $Z\to H+\gamma$ can be expressed in terms of helicity amplitudes
\begin{subequations}\label{eq-gen-rate-helicity-explicit}
\begin{eqnarray}
\Gamma(Z\to \eta_{Q}+\gamma)&=&
\frac{1}{3}\frac{1}{2m_Z}\frac{1}{8\pi}\frac{2|{\bf P}|}{m_Z}\bigg(2| A_{0,1}^{\eta_{Q}}|^2\bigg),\\
\Gamma(Z\to J/\psi(\Upsilon)+\gamma)&=&
\frac{1}{3}\frac{1}{2m_Z}\frac{1}{8\pi}\frac{2|{\bf P}|}{m_Z}\bigg(2| A_{1,1}^{J/\psi(\Upsilon)}|^2
+2| A_{0,1}^{J/\psi(\Upsilon)}|^2\bigg),
\end{eqnarray}
\end{subequations}
where $|{\bf P}|$ denotes the magnitude of the $H$ spatial momentum:
\beq
|{\bf P}|=\frac{\lambda^{1/2}(m_Z^2,m_H^2,0)}{2m_Z}=\frac{m_Z^2-m_H^2}{2m_Z},
\eeq
where $m_H$ refers to the mass of the quarkonium $H$, and the K\"allen function is defined via $\lambda(x,y,z)=x^2+y^2+z^2-2xy-2xz-2yz$.
$A_{\lambda_1, \lambda_2}^{H}$ represents the helicity amplitude of $Z\to H(\lambda_1)+\gamma(\lambda_2)$
with $\lambda_1$ and $\lambda_2$ being the helicities of the $H$ and outgoing photon respectively.
To deduce (\ref{eq-gen-rate-helicity-explicit}), we have applied the parity invariance~\cite{Haber:1994pe}  to relate different helicity amplitude
\bqa\label{eq-helicity-parity-invariance}
A_{0,1}^{\eta_{Q}}=-A_{0,-1}^{\eta_{Q}}, \quad  A_{1,1}^{J/\psi(\Upsilon)}=-A_{1,-1}^{J/\psi(\Upsilon)},  \quad  A_{0,1}^{J/\psi(\Upsilon)}=-A_{0,-1}^{J/\psi(\Upsilon)}.
\eqa
Obviously, there are one independent helicity amplitude for $\eta_Q$ production, and two for $J/\psi$ or $\Upsilon$.

In the limit of $ m_Q \ll m_Z$, the helicity amplitude $A_{\lambda_1,\lambda_2}^H$ satisfies the asymptotic behavior
\bqa\label{eq-helicity-selection-rule}
A_{\lambda_1,\lambda_2}^H \propto r^{1+|\lambda_1|},
\eqa
where $r=m_Q/m_Z$.  In (\ref{eq-helicity-selection-rule}), one power of $r$ originates from the large
momentum transfer which are required for the heavy-quark pair to form the heavy quarkonium with small relative
momentum, and the other powers arise from the helicity selection rule in perturbative QCD~\cite{Chernyak:1980dj,Brodsky:1981kj}.

To obtain the decay width, it is crucial to work out each helicity amplitude, which is the chief task of this work. 
$Z$ boson interacts with quark-antiquark pair through the tree-level weak interaction as
\bqa\label{eq-z-qq}
i\mathcal{L}_{ZQ\bar{Q}}=i\frac{g}{4 c_W}\bar{Q}\gamma^\mu(g_V-g_A\gamma_5)QZ_{\mu},
\eqa
where $g$ is the weak coupling in $SU(2)_L\times U(1)_Y$ electroweak gauge theory, $g_V=1-8s^2_W/3$ and $g_A=1$ for the up-type quark,
and $g_V=-1+4s^2_W/3$ and
$g_A=-1$ for the down-type quark. Here we have defined $s_W\equiv \sin\theta_W$, and $c_W\equiv \cos\theta_W$,
where $\theta_W$ signifies the Weinberg angle.

The $Z$ boson can decay to $\eta_{Q}+\gamma$ through the vectorial interaction, while
decay to $J/\psi(\Upsilon)+\gamma$ through the axial-vectorial interaction. 
For simplicity, it is convenient to explicitly extract the electroweak coupling from the helicity amplitudes:
\begin{subequations}
\begin{eqnarray}\label{eq-helicity-amplitude-1}
A_{\lambda_1,\lambda_2}^{\eta_{Q}}&=&\frac{gg_Vee_Q}{4c_W}\mathcal{A}_{\lambda_1,\lambda_2}^{\eta_{Q}},\\
A_{\lambda_1,\lambda_2}^{J/\psi(\Upsilon)}&=&\frac{gg_Aee_Q}{4c_W}\mathcal{A}_{\lambda_1,\lambda_2}^{J/\psi(\Upsilon)}.
\end{eqnarray}
\end{subequations}


\section{NRQCD computation\label{sec-NRQCD}}
\subsection{The NRQCD factorization}

According to the NRQCD factorization formalism~\cite{Bodwin:1994jh},
the helicity amplitude ${\mathcal A}_{\lambda_1,\lambda_2}^{H}$ can be
factorized into
\beq\label{eq-nrqcd}
\mathcal{A}_{\lambda_1, \lambda_2}^H= \sqrt{2 m_H} \mathcal{C}_{\lambda_1, \lambda_2}^H  \frac{\langle\mathcal{O}\rangle_H}{\sqrt{2N_c}2m_Q}.
\eeq
The nonrelativistically normalized long-distance matrix elements (LDMEs) are
\begin{eqnarray}
\langle \mathcal{O} \rangle_{H}&=&|\langle H|
\psi^{\dagger}\chi |0\rangle|,
\end{eqnarray}
for $\eta_Q$, and
\begin{eqnarray}
\langle \mathcal{O}\rangle_{H}&=&|\langle H|
\psi^{\dagger}(\bm{\sigma}\cdot \bm{\epsilon}_{H}) \chi |0\rangle|,
\end{eqnarray}
for $J/\psi$ or $\Upsilon$, where $\psi^\dagger$ and $\chi$ denote the Pauli spinor fields creating a heavy quark and antiquark in NRQCD respectively, and $\epsilon_{H}$ represents the polarization vector of $J/\psi$ or $\Upsilon$.
The dimensionless SDC $\mathcal{C}_{\lambda_1,\lambda_2}^H$, signifying the perturbative contribution, can be evaluated either 
through the standard matching procedure or the method of region~\cite{Beneke:1997zp}. In this work, we use the latter to compute these
SDCs.
The asymptotic behavior of the helicity  SDCs can be straightforwardly deduced from (\ref{eq-helicity-selection-rule}) and (\ref{eq-nrqcd})
\bqa\label{eq-helicity-scaling-rule-1}
{\mathcal C}_{\lambda_1,\lambda_2}^H\propto r^{|\lambda_1|},
\eqa
by noting that $\langle\mathcal{O}\rangle_H\propto m_Q^{3/2}$.

It is convenient to expand the SDCs in powers of $\alpha_s$
\bqa\label{eq-sdcs-expand}
\mathcal{C}_{\lambda_1,\lambda_2}^H&=&\mathcal{C}_{\lambda_1,\lambda_2}^{H,(0)}\bigg[1+\frac{\alpha_s}{\pi}
\mathcal{C}_{\lambda_1,\lambda_2}^{H,(1)}
+\frac{\alpha_s^2}{\pi^2}\bigg(\frac{\beta_0}{4}\ln\frac{\mu_R^2}{m_Q^2}\mathcal{C}_{\lambda_1,\lambda_2}^{H,(1)}
+\gamma_H \ln\frac{\mu_\Lambda^2}{m_Q^2}\nn\\
&&~~~~~~~~~~~+ \mathcal{C}_{{\rm reg},\lambda_1,\lambda_2}^{H,(2)}+ \mathcal{C}_{{\rm nonreg},\lambda_1,\lambda_2}^{H,(2)}\bigg)\bigg]
+\mathcal{O}(\alpha_s^3),
\eqa
where $\mu_R$ and $\mu_\Lambda$ indicate the renormalization scale and factorization scale respectively,
$\beta_0=(11/3)C_A-(4/3)T_Fn_f$ is the one-loop coefficient of the QCD $\beta$ function,
where $n_f$ is the number of active quark flavors.
The explicit $\ln\mu_R^2$ term is deduced from the renormalization-group invariance.
$\gamma_H$ represents the anomalous dimension associated with the NRQCD
bilinear currents carrying the quantum number $^1S_0$ or $^3S_1$~\cite{Hoang:2006ty}:
\begin{subequations}\label{eq-ano-dim}
\begin{eqnarray}
\gamma_{^1S_0}&=&-\pi^2\bigg(\frac{C_AC_F}{4}+\frac{C_F^2}{2}\bigg),\\
\gamma_{^3S_1}&=&-\pi^2\bigg(\frac{C_AC_F}{4}+\frac{C_F^2}{6}\bigg).
\end{eqnarray}
\end{subequations}
The occurrence of  $\ln\mu_\Lambda^2$ is demanded by the NRQCD factorization. 
Note that the $\gamma_H\ln\mu_\Lambda^2$ terms in (\ref{eq-sdcs-expand}) exactly cancel the $\mu_\Lambda$ dependence of the NRQCD 
matrix element, so that the helicity amplitudes/decay widths are independent of $\mu_\Lambda$.
For convenience, we have classified the two-loop Feynman diagrams into two groups, the `regular' and the `nonregular'. Some representative Feynman diagrams are illustrated in Fig.~\ref{fig-feynman-diagram}. 
Correspondingly,  $\mathcal{C}_{{\rm reg},\lambda_1,\lambda_2}^{H,(2)}$ and $\mathcal{C}_{{\rm nonreg},\lambda_1,\lambda_2}^{H,(2)}$ in (\ref{eq-sdcs-expand}) represent the contributions from the `regular' part and `nonregular' part respectively.

\subsection{The SDCs through $\mathcal{O}(\alpha_s^2)$}

\begin{figure}[htbp]
	\centering
	\includegraphics[width=1\textwidth]{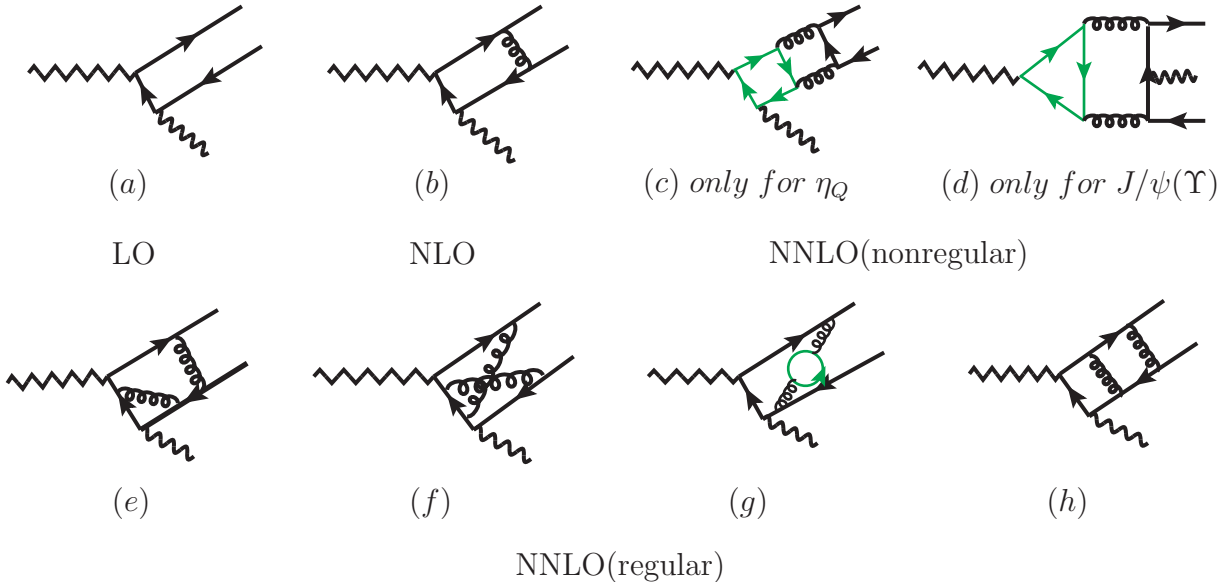}
	\caption{Some representative Feynman diagrams for the process $Z\to H+\gamma$ up to $\mathcal{O}(\alpha_s^2)$.
	\label{fig-feynman-diagram}}
\end{figure}

We briefly outline the calculation. 
We begin with the quark-level process for $Z\to Q\bar{Q}+\gamma$.
The package {\tt FeynArts}~\cite{Hahn:2000kx} is employed to generate the Feynman
diagrams and the corresponding Feynman amplitudes through order $\mathcal{O}(\alpha_s^2)$.
We utilize the spin projectors to enforce $Q\bar{Q}$ in $^1S_0$ for $\eta_Q$ and $^3S_1$ for $J/\psi$ or $\Upsilon$, and
 employ the packages {\tt FeynCalc}~\cite{Shtabovenko:2016sxi} and {\tt FormLink}~\cite{Feng:2012tk} to 
obtain the hadron-level amplitudes order by order in $\alpha_s$.
The helicity amplitudes are evaluated with the aid of the helicity projectors, which are constructed in Appendix.~\ref{appendix-helicity-projectors}.

It is well known that, in dimensional regularization, the
anticommutation relation $\{\gamma^\mu,\gamma_5\}$ and the cyclicity of Dirac
trace cannot be satisfied simultaneously. In practical computation,
the naive-$\gamma_5$ scheme~\cite{Korner:1991sx}, which keeps the anticommutation relation $\{\gamma^\mu,\gamma_5\}$, is frequently applied.
In this scheme, spurious anomaly, which spoils chiral symmetry
and hence gauge invariance, can be avoided. Because
of the lack of the cyclicity of the trace, one must fix a
reading point for a fermion loop with an odd number of $\gamma_5$.
In this work, we will select the vertex of the Z boson as the reading point for the Feynman diagram in Fig.~\ref{fig-feynman-diagram}(d), and the final state quarkonium as the reading point for the other Feynman diagrams.
For more details, we refer the readers to Ref.~\cite{Sang:2022erv}.

At lowest order in $v$, we neglect the relative momentum in $Q\bar{Q}$ pair prior to carrying out the
loop integration, which amounts to directly extracting the SDCs from the hard loop region~\cite{Beneke:1997zp}.
With the aid of the packages {\tt Apart}~\cite{Feng:2012iq} and {\tt FIRE}~\cite{Smirnov:2014hma}, we can reduce the loop integrals into linear combinations of master integrals (MIs).
Finally, we end up with 6 one-loop MIs, and around 320 two-loop MIs, which are evaluated by the powerful package {\tt AMFlow}~\cite{Liu:2022chg,Liu:2017jxz,Liu:2020kpc,Liu:2021wks,Liu:2022mfb}.

After implementing the on-shell renormalization scheme for the heavy quark mass and field strength~\cite{Broadhurst:1991fy,Melnikov:2000zc}, and
the $\overline{\mathrm{MS}}$ renormalization scheme for the QCD coupling, the
UV pole is exactly eliminated, while an uncanceled
single IR pole still remains, which can be factored into the NRQCD LDME, so that the SDC becomes IR finite.

It is straightforward to obtain the LO helicity SDCs:
\begin{subequations}
\begin{eqnarray}
\mathcal{C}_{0,1}^{\eta_{Q},(0)}&=&2\sqrt{6},\\
\mathcal{C}_{1,1}^{J/\psi(\Upsilon),(0)}&=&4\sqrt{6}r  ~~~~~~\mathcal{C}_{0,1}^{J/\psi(\Upsilon),(0)}=-2\sqrt{6}.
\end{eqnarray}
\end{subequations}

The analytic expressions of $\mathcal{C}_{\lambda_1,\lambda_2}^{H,(1)}$ can also be readily obtained.
Instead of presenting the cumbersome expressions, we list their asymptotic expansions in the limit of $r \to 0$:
\begin{subequations}\label{eq-SDCs-1}
\begin{eqnarray}
\mathcal{C}_{0,1}^{\eta_{Q},(1)}&=&\frac{1}{3} (2\ln 2-3) \ln \left(-r^2+i \epsilon\right)-\frac{\pi ^2}{9}-3+\frac{\ln
	^22}{3}+\ln 2, \\
\mathcal{C}_{1,1}^{J/\psi(\Upsilon),(1)}&=&-\frac{2}{3} \ln \left(-r^2+i\epsilon\right)-3-2 \ln 2, \\
\mathcal{C}_{0,1}^{J/\psi(\Upsilon),(1)}&=&\frac{1}{3} (2\ln 2-3) \ln \left(-r^2+i \epsilon\right)-\frac{\pi ^2}{9}-3+\frac{\ln ^2 2}{3}-\frac{\ln 2}{3},
\end{eqnarray}
\end{subequations}
where the real part of  (\ref{eq-SDCs-1}a) is consistent with that in Refs.~\cite{Sang:2009jc,Wang:2013ywc},
and the expression of (\ref{eq-SDCs-1}c) is consistent with that in Ref.~\cite{Wang:2013ywc}.

It becomes much more challenging to deduce the analytical expressions for all the
encountered two-loop MIs.  In this work, we are content with high-precision numerical results.
To perform the numerical computation, we take $m_Z=91.1876$ GeV from the particle data group (PDG)~\cite{ParticleDataGroup:2020ssz},
and the charm quark and bottom quark pole masses to be $m_c = 1.69$ GeV and $m_b=4.80$ GeV, which are converted from the $\overline{\rm MS}$ masses $\overline{m}_c(\overline{m}_c)=1.28$ GeV and $\overline{m}_b(\overline{m}_b)=4.18$ GeV~\cite{ParticleDataGroup:2020ssz} at two-loop level by use of the package {\tt RunDec}~\cite{Chetyrkin:2000yt}.
The numerical values of the various helicity SDCs are tabulated in Table~\ref{tab-sdcs-mc-1.69}. For reference, we explicitly keep 
the $n_L$, $n_c$, and $n_b$ dependence for the SDCs, where $n_L$ denotes the number of light quark flavors, and $n_c=1$ and $n_b=1$ signify the numbers of the charm and bottom quarks respectively. 
The dependence of the theoretical results on the heavy quark mass will be investigated in Sec.~\ref{sec-phen}.
\begin{table}[htbp]\small
	\caption{NRQCD predictions to the various helicity SDCs.
	For simplicity, we define the symbols $f_{1}\equiv\frac{g_V^d}{g_V^u}=-\frac{3-4 S_W^2}{3-8 S_W^2}$, $f_{2}\equiv\frac{g_V^u-g_V^d}{g_V^u}=\frac{6-12 S_W^2}{3-8 S_W^2}$, $\bar{f_{1}}\equiv\frac{g_V^u}{g_V^d}=-\frac{3-8 S_W^2}{3-4 S_W^2}$ and $\bar{f_{2}}\equiv\frac{2g_V^d-2g_V^u}{g_V^d}=\frac{12-24 S_W^2}{3-4 S_W^2}$, where $g_V^u$ and $g_V^d$ correspond to the values of $g_V$ for
	up-type quark and down-type quark respectively.}
	\label{tab-sdcs-mc-1.69}
	\setlength{\tabcolsep}{0.4mm}
	\centering
\resizebox{\textwidth}{!}{
	\begin{tabular}{|c|c|c|c|c|}
		\hline
		{$H$}
		& $(\lambda_1,\lambda_2)$
		& $\mathcal{C}_{\lambda_1,\lambda_2}^{(1)}$
		&$\mathcal{C}_{{\rm reg},\lambda_1,\lambda_2}^{(2)}$
		&$\mathcal{C}_{{\rm nonreg},\lambda_1,\lambda_2}^{(2)}$\\
		\hline
		\multirow{2}*{$\eta_{c}$}
		& \multirow{2}*{(0,1)}
		& \multirow{2}*{$1.035-1.679 i$}
		& $-60.56+27.66 i-(0.88+0.74 i) n_L$
		& $-(2.52 -4.28 i)n_c+(0.70-1.26 i)f_1 n_b$\\
		&
		&
		&$+(0.02-0.75 i) n_c-(0.05+0.77 i) n_b$
		&$-(5.06-0.99 i) f_2$\\
		\hline
		\multirow{4}*{$J/\psi$}
		&  \multirow{2}*{(1,1)}
		& \multirow{2}*{$0.929-2.088 i$}
		& $-59.12+40.50 i-(0.36+1.12 i) n_L$
		& \multirow{2}*{$1.51+1.49 i$}\\
		&
		&
		&$+(0.17-1.12 i)n_c-(0.07+1.15 i)n_b$
		&  \\
		\cline{2-5}
		&  \multirow{2}*{(0,1)}
		& \multirow{2}*{$0.122-1.684 i$}
		& $-50.17+28.13 i-(0.54+0.74 i) n_L$
		& \multirow{2}*{$1.42+1.56 i$}\\
		&
		&
		&$-(0.06+0.75 i)n_c-(0.40+0.77 i)n_b$
		&$$ \\
		\hline
		\multirow{2}*{$\eta_{b}$}
		& \multirow{2}*{(0,1)}
		& \multirow{2}*{$-0.127-1.628 i$}
		& $-46.21+13.95 i-(1.17+0.13 i)n_L$
		& $(7.07-2.89 i)\bar{f_1} n_c-(0.74-4.07 i)n_b$\\
		&
		&
		&$+(0.72-0.14 i) n_c-(0.25+0.16 i) n_b$
		&$-(3.37-0.87 i)\bar{f_2}$\\
		\hline
		\multirow{4}*{$\Upsilon$}
		&  \multirow{2}*{(1,1)}
		& \multirow{2}*{$-0.454-2.065 i$}
		& $-37.76+22.93 i-(0.87+0.37 i) n_L$
		& \multirow{2}*{$-1.47-1.50 i$}\\
		&
		&
		&$+(0.81-0.38 i)n_c-(0.30+0.40 i)n_b$
		&$$  \\
		\cline{2-5}
		&  \multirow{2}*{(0,1)}
		& \multirow{2}*{$-0.985-1.665 i$}
		& $-35.71+15.10 i-(0.85+0.14 i) n_L$
		& \multirow{2}*{$-1.43-1.56 i$}\\
		&
		&
		&$+(0.83-0.15 i) n_c-(0.33+0.17 i) n_b$
		& \\
		\hline
		
	\end{tabular}
}
\end{table}
\newpage
\section{LC factorization for the leading-twist SDCs\label{sec-LL}}

\subsection{The LC factorization~\label{subsec-LL}}
Besides the NRQCD factorization formalism, we can also employ the LC factorization framework
to calculate the decay amplitude for $Z\to H+\gamma$ at the leading twist. By following the spirit of Ref.~\cite{Jia:2008ep},
the LC factorization formula for the SDC is written as
\begin{eqnarray}
\mathcal{C}^H_{0,1}&\equiv & \mathcal{C}^{H,(0)}_{0,1} \mathcal{K}^H=\mathcal{C}^{H,(0)}_{0,1}\int_0^1 d x T_H(x, m_Z,\mu) \hat{\phi}_H(x,m_Q, \mu)+\mathcal{O}(m_Q^2/m_Z^2)\,,
\end{eqnarray}
where the hard-kernel $T_H$ and the leading-twist LC distribution amplitude (LCDA) $\hat\phi_H$ are perturbatively calculable around the scale $m_Z$ and $m_Q$, respectively. Up to $\mathcal{O}(\alpha_s)$, we have the expansions
\begin{subequations}
\begin{eqnarray}
T_H(x, m_Z,\mu)&=&T_H^{(0)}(x)+\frac{\alpha_s(\mu)}{4 \pi} T_H^{(1)}(x,m_Z,\mu)\,,\\
\hat{\phi}_H(x,m_Q,\mu)&=&\hat{\phi}_H^{(0)}(x)+\frac{\alpha_s(\mu)}{4 \pi} \hat{\phi}_H^{(1)}(x,m_Q,\mu)\,.
\end{eqnarray}
\end{subequations}

The explicit hard-kernels and LCDAs up to $\mathcal{O}(\alpha_s)$ are given in \cite{Wang:2013ywc} as \footnote{For convenience, we normalize $\mathcal{K}^H$ equal to 1 at tree-level. Here we adjust the normalizations for hard kernels and  LCDAs given in \cite{Wang:2013ywc} correspondingly.}
\begin{subequations}
\begin{eqnarray}
T_P^{(0)}(x) & = &\frac{1}{4 x\bar x}\,, \\
T_P^{(1)}(x,m_Z, \mu) & =& \frac{C_F}{4 x\bar x}\Bigg\{\left[3+2 x \ln \bar x+2\bar x \ln x\right]\left(\ln \frac{m_Z^2}{\mu^2}-i \pi\right)+x  \ln ^2\bar x+\bar x \ln ^2 x\nonumber\\
&&+(8 \Delta-1)[\bar x \ln \bar x+x \ln x]-9\Bigg\}\,,\\
\hat{\phi}_P^{(0)}\left(x\right) & =&\delta(x-1 / 2)\,, \\
\hat{\phi}_P^{(1)}\left(x, m_Q,\mu\right) & =&C_F \theta(1-2 x)\left\{\left[\left(4 x+\frac{8 x}{1-2 x}\right)\left(\ln \frac{\mu^2}{m_Q^2(1-2 x)^2}-1\right)\right]_{+}\right. \nonumber\\
&&+{\left.\left[\frac{16 x\bar{x}}{(1-2 x)^2}\right]_{++}+\Delta[16 x]_{+}\right\}+(x \leftrightarrow \bar{x}) }+C_F \left(4 \Delta-6\right)\delta(x-1/2)\nn\\
\end{eqnarray}
\end{subequations}
for $H=\eta_{Q}$,
\begin{subequations}
\begin{eqnarray}
T_V^{(0)}(x, m_Z,\mu) & = &\frac{1}{4 x\bar x}\,, \\
T_V^{(1)}(x, m_Z,\mu) & =&\frac{C_F}{4 x\bar x}\Bigg\{\left[3+2 x \ln \bar x+2\bar x \ln x\right]\left(\ln \frac{m_Z^2}{\mu^2}-i \pi\right)+x  \ln ^2\bar x+\bar x \ln ^2 x\nonumber\\
&&-[\bar x \ln \bar x+x \ln x]-9\Bigg\}\,,\\
\hat{\phi}_V^{(0)}\left(x\right) & =&\delta(x-1 / 2)\,, \\
\hat{\phi}_V^{(1)}\left(x, m_Q,\mu\right) & =&C_F \theta(1-2 x)\left\{\left[\left(4 x+\frac{8 x}{1-2 x}\right)\left(\ln \frac{\mu^2}{m_Q^2(1-2 x)^2}-1\right)\right]_{+}\right. \nonumber\\
\quad&&+  {\left.\left[\frac{16 x\bar{x})}{(1-2 x)^2}\right]_{++}-[8 x]_{+}\right\}+(x \leftrightarrow \bar{x}) }-8C_F \delta(x-1/2)\,,
\end{eqnarray}
\end{subequations}
for $H=J/\psi$ and $\Upsilon$. Here $\bar{x}=1-x$ and $\Delta=0$ for the NDR scheme and $\Delta=1$ for the HV scheme.
Note that the terms proportional to $\delta(x-1/2)$ in $\hat\phi_H^{(1)}$ actually contribute to the one-loop corrections to decay constants $f_{P,V}$.

\subsection{The NLL resummation with the ERBL equation~\label{subsec-resum}}
The leading twist LCDAs of quarkonia obey the celebrated ERBL  equation \cite{Lepage:1979zb, Efremov:1979qk}
\begin{eqnarray}
\mu^2 \frac{d}{d \mu^2} \hat\phi_H(x, m_Q,\mu)=\int_0^1 d y V_H\left(x,y ; \alpha_s(\mu)\right) \hat\phi_H(y,m_Q, \mu)\,,\label{eq:ERBLeq}
\end{eqnarray}
with the BL kernel expanded in $\alpha_s$
\begin{eqnarray}
V_H\left(x,y ; \alpha_s(\mu)\right)=\frac{\alpha_s(\mu)}{4 \pi} V^{(0)}_H(x, y)+\left(\frac{\alpha_s(\mu)}{4 \pi}\right)^2 V^{(1)}_H(x, y)+\ldots\,.
\end{eqnarray}
The BL kernel at the lowest order of $\alpha_s$ given in \cite{Lepage:1979zb, Efremov:1979qk} for a pseudoscalar meson is the same as that for a longitudinally polarized vector meson
\begin{eqnarray}
V^{(0)}_P(x,y)=V^{(0)}_V(x,y)\,,	
\end{eqnarray}
while their BL kernels at $\mathcal{O}(\alpha_s^2)$ given in \cite{Sarmadi:1982yg,Dittes:1983dy,Katz:1984gf,Mikhailov:1984ii,Belitsky:1999gu} differ 
from each other by~\cite{Melic:2001wb}
\begin{eqnarray}
 V^{(1)}_P(x,y)-V^{(1)}_V(x,y)=\Delta \left(-8 C_F \left[v^a(x,y)\right]_+\right)\,,
\end{eqnarray}
where $\Delta=0$ for the NDR scheme and $\Delta=1$ for the HV scheme. The explicit expressions of these BL kernels are summarized in Appendix \ref{appendix-BL-kernels}.

The formal solution can be
\begin{eqnarray}
\hat\phi_H(x,m_Q,\mu)&=&\left[U_H(\mu,\mu_0)\ast\hat\phi_H\right](x,m_Q,\mu_0)\,,
\end{eqnarray}
where the evolution kernel is
\begin{eqnarray}
U_H(\mu,\mu_0)&\equiv & \mathrm{P}\exp\left\{\int_{\alpha_s(\mu_0)}^{\alpha_s(\mu)}d\alpha_s\frac{V_H(\alpha_s)}{\beta(\alpha_s)}\right\}\,,
\end{eqnarray}
with $\mathrm{P}$ standing for the ordering on $\alpha_s$ which can be expanded in form of the Dyson series
\begin{eqnarray}
U_H(\mu,\mu_0)&=& 1+\int_{\alpha_s(\mu_0)}^{\alpha_s(\mu)}d\alpha_s\frac{V_H(\alpha_s)}{\beta(\alpha_s)}+\int_{\alpha_s(\mu_0)}^{\alpha_s(\mu)}d\alpha_{s1}\frac{V_H(\alpha_{s1})}{\beta(\alpha_{s1})}\ast\int_{\alpha_s(\mu_0)}^{\alpha_{s1}}d\alpha_{s2}\frac{V_H(\alpha_{s2})}{\beta(\alpha_{s2})}\nonumber\\ &&
+\cdots\,,
\end{eqnarray}
and the ``$\ast$" denotes the appropriate convolution over the light-fractions. Therefore, we have the renormalization group improved SDCs for $Z\to \eta_Q+\gamma$ and $Z\to J/\psi(\Upsilon)+\gamma$ at the leading power of expansion in NRQCD factorization as
\begin{eqnarray}
\mathcal{K}^H&=& \int_0^1 d x T_H\left(x, m_Z, m_Z\right) U_H\left(m_Z, m_Q\right)\hat\phi_H\left(x, m_Q, m_Q\right)+\mathcal{O}\left(m_Q^2 / m_Z^2\right)\,.
\end{eqnarray}

\subsubsection{The truncated NLL resumed SDCs}
The explicit truncation of the perturbative expansion of $U_H$ up to $\mathcal{O}(\alpha_s^2)$ is
\begin{eqnarray}
U_H(\mu,\mu_0)&\approx& 
1+\frac{\alpha_s(\mu)}{4\pi}\ln\frac{\mu^2}{\mu_0^2}V_H^{(0)}+\left(\frac{\alpha_s(\mu)}{4\pi}\right)^2\left(\frac{1}{2}\ln^2\frac{\mu^2}{\mu_0^2}(V_H^{(0)}\ast V_H^{(0)}+\beta_0 V_H^{(0)})+\ln\frac{\mu^2}{\mu_0^2} V_H^{(1)}\right)\,.\nonumber\\
\label{eq:expandedNLL}
\end{eqnarray}
Here we have used the RG evolution of the strong coupling $\alpha_s(\mu)$ at the NLL level
\begin{eqnarray}\label{eq:as-run}
\alpha_s(\mu)&= &\alpha_s(\mu_0)\left[1+\beta_0\frac{\alpha_s(\mu_0)}{4\pi}\ln\frac{\mu^2}{\mu_0^2}+\beta_1\left(\frac{\alpha_s(\mu_0)}{4\pi}\right)^2 \ln\frac{\mu^2}{\mu_0^2}\right]^{-1}\,,
\end{eqnarray}
and its perturbative expansion
\begin{eqnarray}\label{eq:as-run-expand}
	\alpha_s(\mu)&= &\alpha_s(\mu_0)\left[1-\frac{\alpha_s(\mu_0)}{4\pi}\left(\beta_0\ln\frac{\mu^2}{\mu_0^2}\right)+\left(\frac{\alpha_s(\mu_0)}{4\pi}\right)^2 \left(\beta_0^2\ln^2\frac{\mu^2}{\mu_0^2}-\beta_1\ln\frac{\mu^2}{\mu_0^2}\right)+\mathcal{O}(\alpha_s^3)\right]\,.\nonumber\\
\end{eqnarray}

With the useful convolutions listed in Appendix \ref{appendix-convolutions}, we have the truncation of the NLL resumed SDCs up to $\mathcal{O}(\alpha_s^2)$ explicitly
\begin{eqnarray}\label{eq:P-NLL-expand}
\mathcal{K}^{P,\mathrm{NLL}}&=& 1+\frac{\alpha_s(m_Z)}{4\pi}C_F\left[(3-2\ln 2) \left(\ln\frac{m_Z^2}{m_Q^2}-i \pi\right) + \ln^2 2 + 3\ln 2 -\frac{\pi^2}{3} -9
  \right]\nonumber\\
&&+\left(\frac{\alpha_s(m_Z)}{4\pi}\right)^2C_F\Bigg\{\left[C_F\left(\ln^22-8 \ln 2-\frac{\pi^2}{6}+\frac{9}{2}\right)+ \frac{\beta_0}{2}\left(3-2\ln2\right)\right]\ln^2\frac{m_Z^2}{m_Q^2}\nonumber\\
&&
+\Bigg[ C_F\Bigg(\frac{7 }{2}\zeta (3)-\frac{4}{3} \ln^3 2+\frac{5}{3} \pi ^2 \ln 2+6 \ln
   ^22+21 \ln 2-\frac{5 }{3}\pi ^2-\frac{51}{2}\nonumber\\
   &&~~~~-2 i \pi\left(\ln^22-8 \ln 2-\frac{\pi^2}{6}+\frac{9}{2}\right)\Bigg)-C_A\left(\frac{3}{2}\zeta(3)-\frac{4}{3}\ln 2-1\right) \nonumber\\
   &&~~~~- \beta_0 \left(\ln^2 2-\frac{2}{3}\ln 2+\frac{\pi^2}{6}-\frac{1}{2}\right)\Bigg]\ln\frac{m_Z^2}{m_Q^2}\Bigg\}+...	\,,
\end{eqnarray}
for $H=\eta_Q$, and
\begin{eqnarray}\label{eq:V-NLL-expand}
\mathcal{K}^{V,\mathrm{NLL}}
   &=& \mathcal{K}^{P,\mathrm{NLL}}+\frac{\alpha_s(m_Z)}{4\pi}C_F\Bigg[-4 \ln 2\Bigg]\nonumber\\
   &&+\left(\frac{\alpha_s(m_Z)}{4\pi}\right)^2\ln\frac{m_Z^2}{m_Q^2}\Bigg\{C_F^2\Bigg[4\ln^2 2-4\ln 2 \Bigg]-4C_F\beta_0\ln 2\Bigg\}+...\,,
\end{eqnarray}
for $H=J/\Psi(\Upsilon)$. Note that the $\gamma_5$-scheme dependences of the hard-kernel $T_P$ and LCDA $\hat \phi_P$ cancel with each other eventually in the above expansion of $\mathcal{K}^{P,\rm NLL}$ as it should.

\subsubsection{The NLL resummation to all orders with the Gegenbauer polynomial expansion}
The resummation of the LL to all orders of $\alpha_s$ are commonly done with the assists of the Gegenbauer polynomials by noticing such polynomials are the eigenfuctions of the BL kernel at the lowest order of $\alpha_s$. The NLL resummation to all orders of $\alpha_s$ can be also done in a similar way by considering the nondiagonal part in $V^{(1)}_H(x,y)$.

The Gegenbauer expansion of the LCDA at twist-2 is
\begin{eqnarray}
\hat\phi_H(x, m_Q,\mu) \equiv \sum_{n=0}^{\infty} \hat\phi_{H,n}(\mu) x(1-x) C_n^{(3 / 2)}(2 x-1)\,,	
\end{eqnarray}
with $C^{(3/2)}_n(x)$ being the order $3/2$ Gegenbauer polynomials, and the Gegenbauer moments
\begin{eqnarray}
\hat\phi_{H,n}(\mu)=\frac{4(2 n+3)}{(n+1)(n+2)}\int_0^1 d x C_n^{(3 / 2)}(2 x-1) \hat\phi_H(x, m_Q, \mu) \,.	
\end{eqnarray}
$\hat\phi_{H,n}(\mu)$ can have the similar perturbative expansion in $\alpha_s$ around the scale $m_Q$ as $\hat\phi_H(x,m_Q,\mu)$.

Solving the ERBL equation in the Gegenbauer moments space, we have formally
\begin{eqnarray}
\hat\phi_{H,n}(\mu)=\sum_{k=0}^n U_{n, k}^H\left(\mu, \mu_0\right) \hat\phi_{H,k}\left(\mu_0\right)\,,
\end{eqnarray}
in which the matrix-elements of the NLL evolution kernel in the Gegenbauer moments space $U_{n,k}^H$ are given in Ref.~\cite{Agaev:2010aq}.

Similarly, we have the Gegenbauer expansion for the hard-kernels
\begin{eqnarray}
T_H(x, m_Z, \mu)=\sum_{n=0}^{\infty} \frac{4(2 n+3)}{(n+1)(n+2)} T_{H,n}(\mu) C_n^{(3 / 2)}(2 x-1)\,,
\end{eqnarray}
with
\begin{eqnarray}
T_{H,n}(\mu)=\int_0^1 d x x(1-x) C_n^{(3 / 2)}(2 x-1) T_H(x, m_Z,\mu)\,.	
\end{eqnarray}
$T_{H,n}(\mu)$ can have the similar perturbative expansion in $\alpha_s$ around the scale $m_Z$ as $T_H(x,m_Z,\mu)$.

Hence, we can get
\begin{eqnarray}
\mathcal{K}^H=\sum_{n=0}^{\infty}\sum_{k=0}^\infty T_{H,n}(m_Z) U^H_{n,k}(m_Z,m_Q)\hat\phi_{H,k}(m_Q)\,.
\end{eqnarray}
Specifically, 
we have
\begin{eqnarray}\label{eq:NLL-resummation-0}
\begin{aligned}
\mathcal{K}^H=  \mathcal{K}^{H(0,0)}+\frac{\alpha_s(m_Z)}{4 \pi} \mathcal{K}^{H(1,0)}+\frac{\alpha_s\left(m_Q\right)}{4 \pi} \mathcal{K}^{H(0,1)}
\end{aligned}	\,,
\end{eqnarray}
where
\begin{eqnarray}\label{eq:NLL-resummation}
\mathcal{K}^{H(i, j)}=\sum_{n=0}^{\infty}\sum_{k=0}^\infty  T_{H,n}^{(i)}(\mu) U^H_{n k}\left(\mu, \mu_0\right) \hat\phi_{H,k}^{(j)}\left(\mu_0\right)\,.
\end{eqnarray}
It is worth mentioning that $\mathcal{K}^{P}$ is indeed independent of the $\gamma_5$-scheme used in calculations of the hard-kernel $T_P$ and LCDA $\hat \phi_P$ as well as $V^{(1)}_P$, as the authors of Ref.~\cite{Chung:2019ota} indicated.

\subsection{The SDCs by combining the NRQCD prediction and the LC resummation}

We proceed to compute the leading-twist SDCs by combining the NRQCD prediction and the LC resummation.
To avoid double counting, one should subtract $\alpha_s^n \ln^n r$ terms from the NRQCD prediction for the LL resummation, 
and subtract both the $\alpha_s^n \ln^n r$ and $\alpha_s^{n+1} \ln^{n} r$ terms for the NLL resummation.

It is convenient to introduce the following symbols
\begin{subequations}
\begin{eqnarray}
\mathcal{C}_{0,1}^{H,\rm LL}&\equiv &\mathcal{C}_{0,1}^{H,(0)}\mathcal{K}^{H(0, 0)}\,,\\
\mathcal{C}_{0,1}^{H,\rm NLL}&\equiv &\mathcal{C}_{0,1}^{H,(0)}\mathcal{K}^{H}\,,\\
{\widetilde{\mathcal{C}}}_{0,1}^{H,\rm LL}&\equiv &\mathcal{C}_{0,1}^{H,(0)}\mathcal{K}^{H,\rm LL}\,,\\
{\widetilde{\mathcal{C}}}_{0,1}^{H,\rm NLL}&\equiv &\mathcal{C}_{0,1}^{H,(0)}\mathcal{K}^{H,\rm NLL}\,,
\end{eqnarray}
\end{subequations}
where $\mathcal{K}^{H,\rm NLL}$ can be found in (\ref{eq:P-NLL-expand}) and (\ref{eq:V-NLL-expand}), and $\mathcal{K}^{H,\rm LL}$
signify the sum of $\alpha_s^n \ln^n r$ terms in $\mathcal{K}^{H,\rm NLL}$.
Thus, we formally have
\begin{subequations}
\bqa\label{eq-NNLO-LL}
\mathcal{C}_{0,1}^{H,{\rm LO+(N)LL}}&=&\mathcal{C}_{0,1}^{H,\rm LO}-
{\widetilde{\mathcal{C}}}_{0,1}^{H,{\rm (N)LL}}\big|_{\alpha_s^0}+\mathcal{C}_{0,1}^{H, \rm (N)LL},\\
\mathcal{C}_{0,1}^{H,{\rm NLO+(N)LL}}&=&\mathcal{C}_{0,1}^{H,\rm NLO}-
{\widetilde{\mathcal{C}}}_{0,1}^{H,{\rm (N)LL}}\big|_{\alpha_s^1}+\mathcal{C}_{0,1}^{H, \rm (N)LL},\\
\mathcal{C}_{0,1}^{H,{\rm NNLO+(N)LL}}&=&\mathcal{C}_{0,1}^{H,\rm NNLO}-
{\widetilde{\mathcal{C}}}_{0,1}^{H,{\rm (N)LL}}\big|_{\alpha_s^2}+\mathcal{C}_{0,1}^{H, \rm (N)LL},
\eqa
\end{subequations}
where the superscripts `LO', `NLO' and `NNLO' indicate the NRQCD SDCs accurate up to $\mathcal{O}(\alpha_s^0)$, $\mathcal{O}(\alpha_s^1)$ and $\mathcal{O}(\alpha_s^2)$ respectively, and ${\widetilde{\mathcal{C}}}_{0,1}^{H,{\rm (N)LL}}\big|_{\alpha_s^n}$ indicates the truncated  ${\widetilde{\mathcal{C}}}_{0,1}^{H,{\rm (N)LL}}$ up to $\mathcal{O}(\alpha_s^n)$.

With $m_c=1.69$ GeV and $m_b=4.8$ GeV, we enumerate the squared SDCs at various levels of accuracy in Table~\ref{tab-SDCs-squared}. 
Note that the strong coupling constant $\alpha_s(m_Q)$ is evaluated through the running formula (\ref{eq:as-run}). To accelerate the convergence, 
we employ the Abel-Pad$\acute{\rm e}$ approach~\cite{Bodwin:2016edd} to sum the series in (\ref{eq:NLL-resummation}).

\begin{table}[!htbp]
	\caption{Squared leading-twist SDCs $|\mathcal{C}_{0, 1}^{H}|^2$ at various levels of accuracy. We take $\mu_R=m_Z$ and $\mu_\Lambda=1$ GeV.
	}
	\label{tab-SDCs-squared}
	\setlength{\tabcolsep}{7pt}
	\renewcommand{\arraystretch}{1.3}
	\resizebox{\textwidth}{!}{
	\centering
	\begin{tabular}{|c|c|c|c|c|c|c|c|c|c|}
		\hline
		{$H$}
		&LO
		&LO+LL
        &LO+NLL
		&NLO
		&NLO+LL
        &NLO+NLL
		&NNLO
		&NNLO+LL
        &NNLO+NLL\\
		\hline
		$\eta_c$
        & 24.0
		& 35.0
		& 23.1
		& 26.0
		& 28.3
		& 23.1
		& 22.8
        & 23.9
        & 19.9 \\
		\hline
		$J/\psi$
        & 24.0
		& 35.0
		& 19.0
		& 24.3
		& 26.6
		& 19.0
		& 22.0
        & 23.0
        & 17.6 \\
		\hline
		$\eta_b$
        & 24.0
		& 31.4
		& 22.6
		& 23.9
		& 25.0
		& 22.5
		& 23.4
        & 23.9
        & 22.2\\
		\hline
		$\Upsilon$
        & 24.0
		& 31.4
		& 19.8
		& 22.3
		& 23.5
		& 19.8
		& 21.7
        & 22.2
        & 19.9\\
		\hline
	\end{tabular}
}
\end{table}

From Table~\ref{tab-SDCs-squared}, we find that the LL resummation can significantly improve the LO NRQCD prediction,
however only slightly alter the higher order predictions. It can be explained by that some dominant $\alpha_s^n\ln^n r$
contributions have already been included in the higher order NRQCD SDCs, i.e.,  the $\mathcal{O}(\alpha_s\ln r)$ contribution has 
been included in the $\mathcal{O}(\alpha_s)$ NRQCD SDC, and the $\mathcal{O}(\alpha^2_s\ln^2 r)$ contribution have been 
included in the $\mathcal{O}(\alpha_s^2)$ NRQCD SDC.

To be contrary,  the effect of the NLL resummation does not become weaker as the perturbative order increases, e.g., the difference between `NNLO+NLL' and 
`NNLO' is as large as that between `NLO+NLL' and 
`NLO' . 
Furthermore, unlike the case for LL resummation,  we find the difference between the NLL resummation $\mathcal{K}^H$ and its truncated expansion $\mathcal{K}^{H,\rm NLL}$ actually does not decrease from one-loop to two-loop accuracy.
The crucial reason is that we evaluate the value of $\alpha_s(m_Q)$ in (\ref{eq:NLL-resummation-0}) by the running formula (\ref{eq:as-run}), while obtain 
its truncated expression $\mathcal{K}^{H,\rm NLL}$ by employing (\ref{eq:as-run-expand}). The values of $\alpha_s(m_Q)$ are quite different when using (\ref{eq:as-run}) and (\ref{eq:as-run-expand}), for example $\alpha_s(m_c)=0.297$ by (\ref{eq:as-run}) and $\alpha_s(m_c)=0.228$ by (\ref{eq:as-run-expand}).~\footnote{By taking $m_c=1.69$ GeV, $m_Z=91.1876$ GeV, and $\alpha_s(m_Z)=0.1181$, we obtain the values of $\mathcal{K}^{J/\psi(0,0)}$, $\mathcal{K}^{J/\psi(1,0)}$ and $\mathcal{K}^{J/\psi(0,1)}$ to be $1.217$, $-10.555-4.227 i$ and $-9.673$, respectively.
As a comparison, if expanding $\mathcal{K}^{J/\psi}$ in powers of $\alpha_s$, and truncating $\mathcal{K}^{J/\psi(0,0)}$ to $\mathcal{O}(\alpha_s^2)$, $\mathcal{K}^{J/\psi(1,0)}$ and $\mathcal{K}^{J/\psi(0,1)}$ to $\mathcal{O}(\alpha_s)$, we obtain the values of the three
quantities to be $1.203$, $-10.658-4.909 i$, and $-8.657$ respectively. }

Since contribution from the NLL resummation is considerable even at two-loop order, particularly for $J/\psi$ production, the NLL 
resummation is important to improve the theoretical predictions.

\section{phenomenology\label{sec-phen}}

In phenomenological analysis, we take $s_W^2=0.231$, $m_Z=91.1876$ GeV, the total decay width of the $Z$ boson $\Gamma_Z=2.4952$ GeV~\cite{ParticleDataGroup:2020ssz}, and fix the running QED coupling $\alpha(m_Z)=1/128.943$~\cite{Sun:2016bel}.
The default value of $\mu_R$ is chosen $\mu_R=m_Z/\sqrt{2}$ and we have varied $\mu_R$ from $m_Z/2$ to $m_Z$ to estimate the theoretical uncertainties
in computing the NNLO perturbative corrections.  In addition, 
we approximate the NRQCD LDMEs at $\mu_\Lambda=1$ GeV by the Schr\"{o}dinger radial wave function at the origin
\begin{subequations}
\begin{eqnarray}\label{eq-ldme-wave-funciton}
|\langle \mathcal{O} \rangle_{\eta_c}|^2&\approx&|\langle \mathcal{O} \rangle_{J/\psi}|^2 \approx \frac{N_c}{2\pi} |R_{1S,c\bar{c}}(0)|^2
=\frac{N_c}{2\pi}\times 0.81\, {\rm GeV^{3}},\\
|\langle \mathcal{O} \rangle_{\eta_b}|^2&\approx&|\langle \mathcal{O} \rangle_{\Upsilon}|^2 \approx \frac{N_c}{2\pi} |R_{1S,b\bar{b}}(0)|^2
=\frac{N_c}{2\pi}\times 6.477\, {\rm GeV^{3}},
\end{eqnarray}
\end{subequations}
where the radial wave functions at the origin are evaluated from Buchm\"uller-Tye (BT) potential model~\cite{Eichten:1995ch}.

By taking the heavy quark pole masses $m_c=1.69$ GeV and $m_b=4.80$ GeV,  we tabulate the unpolarized decay widths and branching fractions for $Z\to H+\gamma$  at various levels of accuracy in Table~\ref{tab:decay-width-ratio}.  
Since the axial-vectorial interaction of $Zc\bar{c}$ is roughly three times larger than the vectorial interaction, 
the branching fraction for $Z\to J/\psi+\gamma$ is much larger than that for $Z\to \eta_c+\gamma$.
In addition, the $\mathcal{O}(\alpha_s)$ corrections are negligible for $J/\psi$ and $\eta_b$ production, while can reach 10\% for the other two channels.  
The $\mathcal{O}(\alpha_s^2)$ corrections are moderate for the charmonium production, however are small for the bottomonium production. 
Moreover, we find the NLL resummation can considerably alter the NRQCD predictions, particularly for $Z\to J/\psi+\gamma$.
It is worth noting that the uncertainty from the renormalization scale $\mu_R$ is inconsiderable. To be honest, 
we must emphasize that different choice of the values of the LDMEs~\cite{Eichten:1995ch,Bodwin:2001mk,Bodwin:1996tg,Bodwin:2007fz,Chung:2010vz,Chung:2020zqc} may largely affect the theoretical predictions.

\begin{table}
	\caption{\label{tab:decay-width-ratio}Unpolarized decay widths and branching fractions for $Z\to H+\gamma$ at various levels of accuracy.  For comparison, the LC predictions from Ref.~\cite{Luchinsky:2017jab} and Ref.~\cite{Grossman:2015cak} are also listed in the fifth column and sixth column respectively. The two uncertainties in the fifth column arise from uncertainties of the leptonic decay constant of charmonium and the LCDA parameters respectively. 
The three uncertainties in the sixth column originate from the factorization scale dependence, the quarkonium decay constants, and the  LCDA parameters, respectively.}
	\setlength{\tabcolsep}{0pt}
	\renewcommand{\arraystretch}{1.3}
	\begin{center}
		\begin{tabular}{ccc ccc ccc ccc ccc ccc}
			\hline
		\ \ Channel \ \
			&&&\ \ Order\ \
			&&&\ \ $\Gamma_{\rm total}(\rm eV)$\ \
			&&&\ \ $\rm Br(\times 10^{-9})$ \ \
			&&&\ $\rm Br(\times 10^{-9})$~\cite{Luchinsky:2017jab}\ \ \ \
			&&&\ $\rm Br(\times 10^{-9})$~\cite{Grossman:2015cak}\ \\
			\hline
			\ \ \multirow{4}{*}{$Z \to \eta_c+\gamma$} \ \
			&&& LO
			&&& $29.1$
			&&& $11.7$
			&&&\multirow{4}{*}{$9.4\pm 1\pm 0.1$}
			&&&\multirow{4}{*}{$-$} \\
			&&& NLO
			&&& $31.7$
			&&& $12.7$\\
			&&& NNLO
			&&& $27.3^{+0.4}_{-0.5}$
			&&&  $10.9^{+0.2}_{-0.2}$\\
			&&& NNLO+NLL
			&&& $23.8^{+0.4}_{-0.5}$
			&&& $9.5^{+0.2}_{-0.2}$ \\
			\hline
			\multirow{4}{*}{$Z \to J/\psi+\gamma$}
			&&& LO
			&&& $197.7$
			&&& $79.2$
			&&&\multirow{4}{*}{$88\pm 9\pm 0.9$}
			&&&\multirow{4}{*}{$80.2_{-1.5-2.0-3.6}^{+1.4+2.0+3.9}$}\\
			&&& NLO
			&&& $200.6$
			&&& $80.4$\\
			&&& NNLO
			&&& $178.8^{+2.1}_{-2.5}$
			&&& $71.6^{+0.9}_{-1.0}$ \\
			&&& NNLO+NLL
			&&& $143.5^{+1.9}_{-2.3}$
			&&& $57.5^{+0.8}_{-0.9}$   \\
			\hline
			\multirow{4}{*}{$Z \to \eta_b+\gamma$}
			&&& LO
			&&& $65.9$
			&&& $26.4$
			&&&\multirow{4}{*}{$-$}
			&&&\multirow{4}{*}{$-$}\\
			&&& NLO
			&&& $65.5$
			&&& $26.3$ \\
			&&& NNLO
			&&& $64.0^{+0.2}_{-0.2}$
			&&& $25.6^{+0.1}_{-0.1}$ \\
			&&& NNLO+NLL
			&&& $60.7^{+0.2}_{-0.2}$
			&&& $24.3^{+0.1}_{-0.1}$ \\
			\hline
			\multirow{4}{*}{$Z \to \Upsilon+\gamma$}
			&&& LO
			&&& $139.2$
			&&& $55.8$
			&&&\multirow{4}{*}{$-$}
			&&&\multirow{4}{*}{$53.9_{-1.0-0.8-0.8}^{+1.0+0.8+1.1}$}\\	
			&&& NLO
			&&& $129.2$
			&&& $51.8$\\
			&&& NNLO
			&&& $125.6^{+0.4}_{-0.4}$
			&&& $50.3^{+0.2}_{-0.2}$ \\
			&&& NNLO+NLL
			&&& $115.4^{+0.4}_{-0.4}$
			&&& $46.3^{+0.2}_{-0.2}$ \\
			\hline
		\end{tabular}
	\end{center}
	\label{Table-1}
\end{table}

It is intriguing to compare our theoretical prediction on the branching fraction with that from the LC models in literature. In Table~\ref{tab:decay-width-ratio}, we 
list the LC predictions from Ref.~\cite{Luchinsky:2017jab} and Ref.~\cite{Grossman:2015cak} in the fifth column and 
sixth column respectively.  In Ref.~\cite{Luchinsky:2017jab}, the authors take the same LCDA for $\eta_c$ and $J/\psi$, which was obtained at $\mu=m_c$ 
in Ref.~\cite{Braguta:2006wr} inspired by the QCD sum rule.  In Ref.~\cite{Grossman:2015cak},  the authors assumed the LCDA for $J/\psi$ and $\Upsilon$ 
at $\mu=1$ GeV  to be of Gaussian form.  Different LCDA corresponds to  the different internal quark motion assumption.
In spite of having very different nonperturbative inputs, we find our `NNLO+NLL' prediction for $\eta_c$ production is consistent with the result from Ref.~\cite{Luchinsky:2017jab}, and our prediction for $\Upsilon$ production roughly agrees with the result from Ref.~\cite{Grossman:2015cak}.
On the other hand, our prediction for $J/\psi$ production is a bit smaller than the values from both references. The distinct LCDA input and 
the considerable NLL resummation effect mainly account for the difference. 
There is no doubt that the ambiguity in choosing LCDA can cause large uncertainties.

\begin{figure}[htbp]
	\centering
	\includegraphics*[scale=0.6]{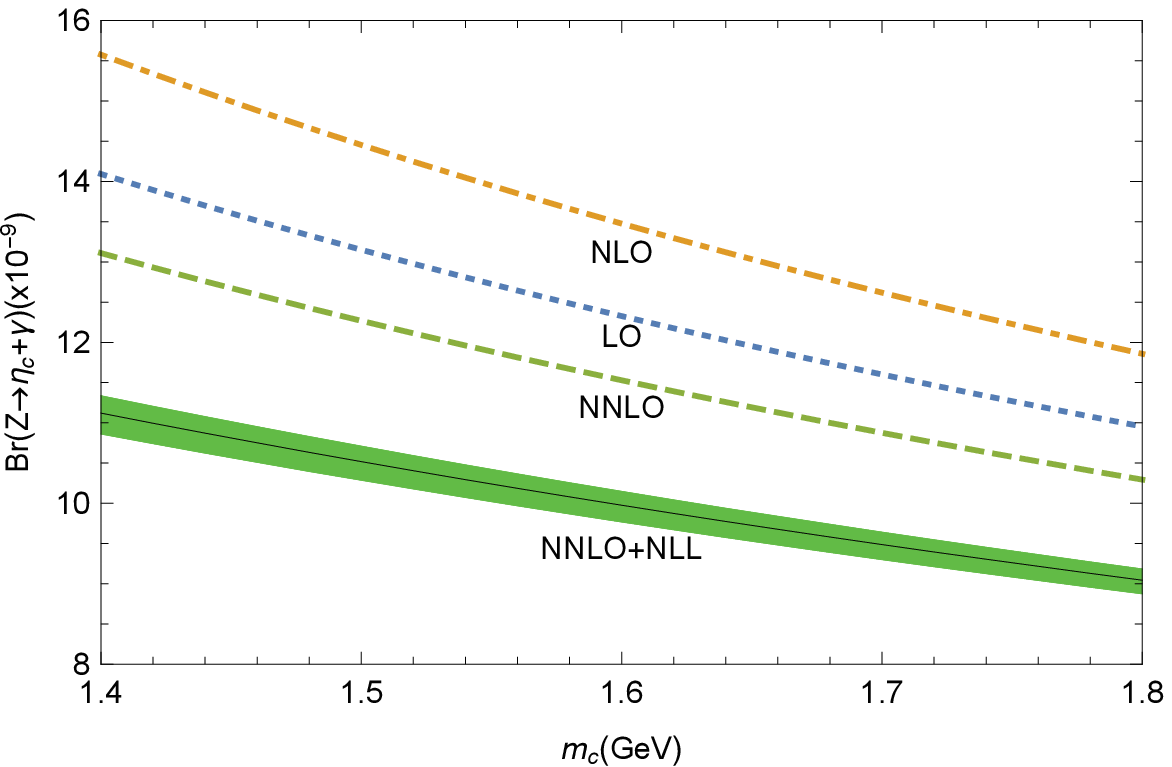}
    \includegraphics*[scale=0.6]{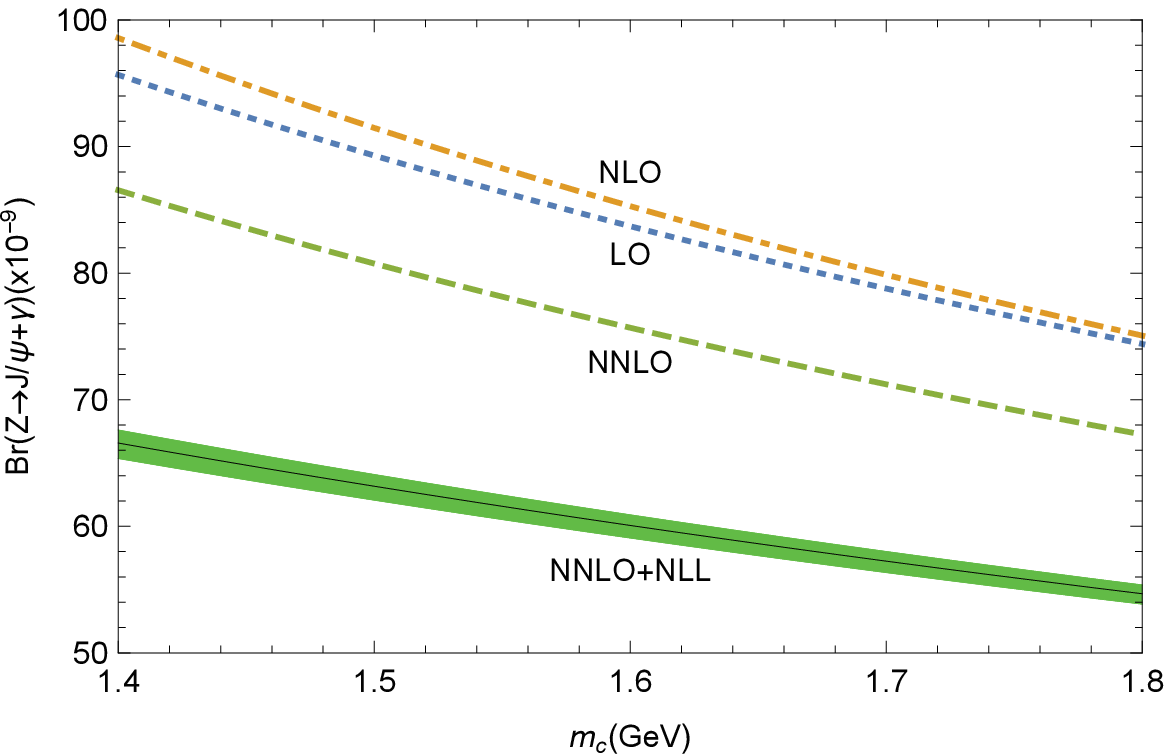}
    \includegraphics*[scale=0.6]{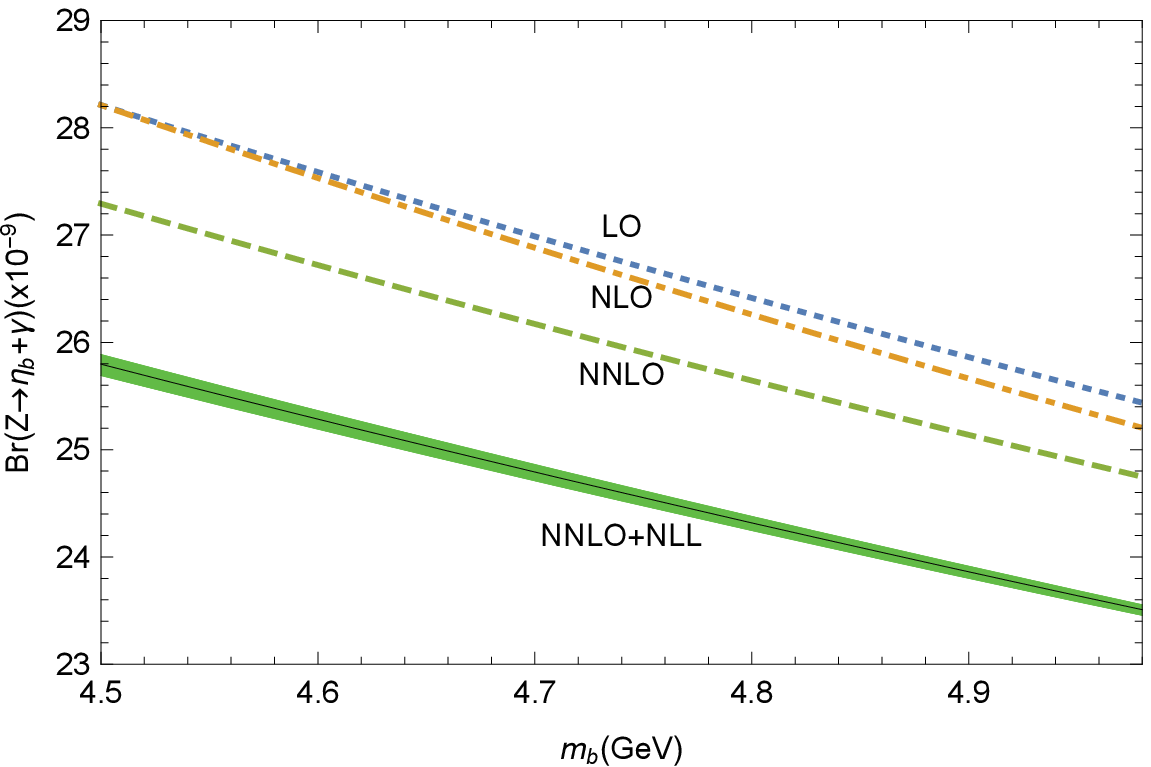}
    \includegraphics*[scale=0.6]{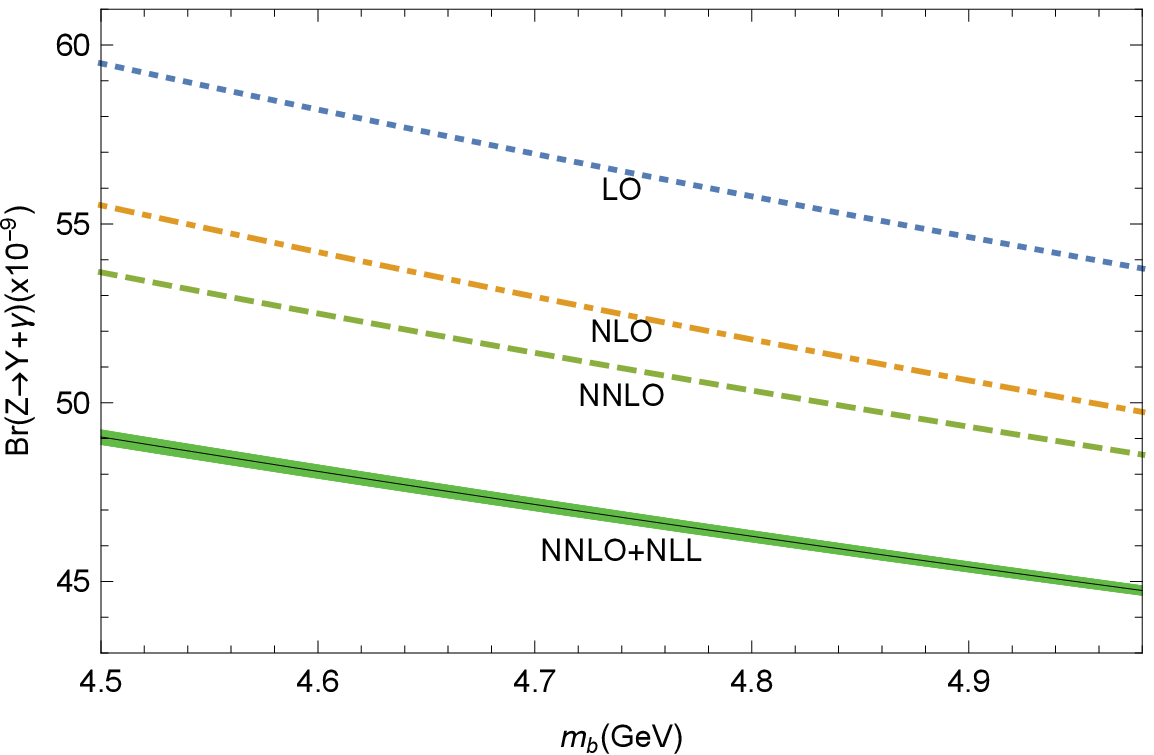}
	\caption{Branching fraction for $Z\to H+\gamma$ as a function of $m_Q$. The green band denotes the uncertainty  from $\mu_R$.
	\label{fig-branching-ratio}}
\end{figure}

We proceed to investigate the dependence of the theoretical results on the heavy quark mass.
In Fig.~\ref{fig-branching-ratio}, we plot the branching fraction for $Z\to H+\gamma$ as a function of $m_Q$ at various levels of accuracy. We find 
that the branching fraction monotonically decreases as $m_Q$ increases.

Finally, we estimate signal events at the future super Z factories, such as the $Z$-factory mode in CEPC, where the $Z$ boson yield can reach $7\times 10^{11}$~\cite{CEPCStudyGroup:2018ghi}. It is expected that there will be around $5\times 10^4$ $J/\psi$ or $\Upsilon$ and $10^4$ $\eta_Q$ events produced through $Z\to H+\gamma$. The $J/\psi$ or $\Upsilon$ can be reconstructed through their leptonic decays, thus providing several thousands of $\gamma \ell\ell$ events. It is promising to search these two channels. However, due to the lack of a clean decay mode for $\eta_Q$, the experimental measurement of $Z\to \eta_Q+\gamma$ would be quite challenging.

\section{summary~\label{sec-summary}}

In summary, we have computed the $\mathcal{O}(\alpha_s)$ and $\mathcal{O}(\alpha_s^2)$ corrections to the helicity amplitudes and decay widths for $Z\to H+\gamma$ by applying
the NRQCD factorization approach. It is found that the $\mathcal{O}(\alpha_s)$ corrections are moderate for $\eta_c$ and $\Upsilon$ productions, 
however tiny for $J/\psi$ and $\eta_b$ productions. The $\mathcal{O}(\alpha_s^2)$ corrections are considerable for charmonium production, while small for bottomonium production. In addition, we find that the branching fraction for $J/\psi$ production is much larger than that for $\eta_c$ production. 

The NLL of $m_Z^2/m_Q^2$ in the leading-twist SDCs are resummed to all orders of $\alpha_s$ by employing the celebrated ERBL equation. 
We find that the NLL resummation can considerably alter the NRQCD prediction, particularly for $J/\psi$ production, so that 
the NLL resummation is important to improve the theoretical prediction. 
We also find the branch fraction for $Z\to H+\gamma$ monotonically decreases as $m_Q$ increases.
In addition, we compare our predictions on the branching fractions with these from the LC models in literature. We find our prediction for $\eta_c$ production 
is consistent with Ref.~\cite{Luchinsky:2017jab}, our prediction for $\Upsilon$ production is slightly smaller than Ref.~\cite{Grossman:2015cak}, and our prediction for $J/\psi$ is a bit smaller than
Ref.~\cite{Grossman:2015cak}. 

It is expected that there will be around $5\times 10^4$ $J/\psi$ or $\Upsilon$ events produced through $Z$ boson radiative decay at 
the future super Z factories.  Therefore it seems that the observation prospects of $Z\to H+\gamma$
are promising in the future.

\begin{acknowledgments}
The authors wish to thank the high performance computing platform of School of Physical Science and Technology of Southwest University for the computational resources support.
The work of W.-L. S. is supported by the National Natural Science Foundation
of China under Grants No. 11975187.
The work of D.~Y. is supported in part by the National Natural Science Foundation of China under Grants Nos.~12235008 and  11635009.
The work of Y.-D. Z. is supported by the National Natural Science Foundation
of China under Grant Nos.~12135006 and 12075097, as well as by the Fundamental Research Funds for the Central Universities under Grant Nos.~CCNU20TS007 and CCNU22LJ004.
This work was also supported in part by the Natural Science Foundation of China under Grant No.11847301 and by the Fundamental Research Funds for the Central Universities under Grant No. 2019CDJDWL0005.

\end{acknowledgments}

\section*{Appendix}
\appendix
\section{Construction of helicity projectors\label{appendix-helicity-projectors}}

In this appendix, we present the various helicity projectors $\mathcal{P}^{(H)}_{\lambda_1,\lambda_2}$ used to compute the helicity amplitudes 
in Sec.~\ref{sec-NRQCD}. The technique in the following is similar to that in Refs.~\cite{Xu:2012uh,Zhang:2021ted}.

For convenience, we introduce two auxiliary longitudinal vectors,
\begin{subequations}\label{eq-auxiliary}
	\begin{eqnarray}
	L_Z^\mu &=& \frac{1}{|\mathbf{P}|} \bigg(P^\mu-\frac{Q\cdot P}{m_Z^2}Q^\mu\bigg),\\
	L_{J/\psi(\Upsilon)}^{\mu}&=&\frac{1}{|\mathbf{P}|} \bigg(\frac{P\cdot Q}{m_Z m_{J/\psi(\Upsilon)}}P^\mu-\frac{m_{J/\psi(\Upsilon)}}{m_{Z}}Q^\mu\bigg),
	\end{eqnarray}
\end{subequations}
where $P$ and $Q$ denote the momenta of the $H$ meson and $Z$ boson, respectively.
The longitudinal vectors satisfy $L_{Z}^\mu Q_\mu=L_{J/\psi(\Upsilon)}^\mu P_\mu=0$.

We present all the 3 helicity projectors
\begin{subequations}
	\begin{eqnarray}
	\mathcal{P}_{0,1}^{(\eta_Q)\mu \nu}&=&\frac{i}{2m_{{Z}} |\mathbf{P}|}\epsilon^{\mu\nu\rho \sigma}
	Q_{\rho} P_{\sigma},\\
	\mathcal{P}_{1,1}^{(J/\psi(\Upsilon))\mu \nu \alpha}&=&\frac{i}{2m_{{Z}} |\mathbf{P}|} L_{Z}^{\mu}\epsilon^{\nu\alpha\rho \sigma}
	Q_{\rho} P_{\sigma},\\
	\mathcal{P}_{0,1}^{(J/\psi(\Upsilon))\mu \nu \alpha}&=&\frac{i}{2m_{Z} |\mathbf{P}|} L_{J/\psi(\Upsilon)}^{\alpha}\epsilon^{\mu\nu\rho \sigma}
	Q_{\rho} P_{\sigma}.
	\end{eqnarray}

If expressing the amplitudes of $Z\to H(\lambda_1)+\gamma(\lambda_2)$ as
	\begin{equation}
	\begin{aligned}
	\mathcal{A}^{(\eta_Q)}&=\mathcal{A}_{\mu \nu}^{(\eta_Q)} \epsilon_Z^\mu \epsilon_\gamma^{* \nu}, \\
	\mathcal{A}^{(J/\psi(\Upsilon))}&=\mathcal{A}_{\mu \nu \alpha}^{\left(J/\psi(\Upsilon)\right)} \epsilon_Z^\mu \epsilon_\gamma^{* \nu} \epsilon_{J/\psi(\Upsilon)}^{* \alpha},
	\end{aligned}
	\end{equation}
we can obtain the helicity amplitude through
	\begin{equation}
	\begin{aligned}
	\mathcal{A}^{(\eta_Q)}_{0,1}&=\mathcal{P}_{0,1}^{(\eta_Q)\mu \nu}\mathcal{A}_{\mu \nu}^{(\eta_Q)}, \\
	\mathcal{A}^{(J/\psi(\Upsilon))}_{0,1}&=	\mathcal{P}_{0,1}^{(J/\psi(\Upsilon))\mu \nu \alpha}\mathcal{A}_{\mu \nu \alpha}^{(J/\psi(\Upsilon))},\\
	\mathcal{A}^{(J/\psi(\Upsilon))}_{1,1}&=	\mathcal{P}_{1,1}^{(J/\psi(\Upsilon))\mu \nu \alpha}\mathcal{A}_{\mu \nu \alpha}^{(J/\psi(\Upsilon))}.
	\end{aligned}
	\end{equation}
\end{subequations}

\section{The explicit expressions for the BL kernels \label{appendix-BL-kernels}}
The BL kernels at the one- and two-loop level have been given in~\cite{Efremov:1979qk, Lepage:1980fj} and~\cite{Sarmadi:1982yg,Dittes:1983dy,Katz:1984gf,Mikhailov:1984ii,Belitsky:1999gu}, respectively. For the nonsinglet evolution up to $\mathcal{O}(\alpha_s^2)$, we have the BL kernels explicitly \cite{Belitsky:1999gu}
\begin{subequations}
\begin{eqnarray}
V^{(0)}_P(x,y)&=&V^{(0)}_V(x,y)=2 C_F\left[v(x,y)\right]_+\,,\\
V^{(1)}_P(x,y)&=&V^{(1)}_V(x,y)-8\Delta C_F\beta_0 \left[v^a(x,y))\right]_+\,,\\
V^{(1)}_V(x,y)&=& 4 C_F\left[C_F V_F(x,y)-\frac{\beta_0}{2} V_\beta(x,y)-\left(C_F-\frac{C_A}{2}\right) V_G(x,y)\right]_+\,,
\end{eqnarray}
\end{subequations}
where
\begin{eqnarray}
v(x,y)&=&f(x,y)\theta(y-x)+f(\bar x,\bar y)\theta(x-y)\,,
~~\text{with} ~~~
f(x,y)\equiv \frac{x}{y}\left(1+\frac{1}{y-x}\right)\,,
\end{eqnarray}
and
\begin{subequations}
\begin{eqnarray}
V_F(x, y) &= &\theta(y-x)\left\{\left(\frac{4}{3}-2 \zeta(2)\right) f+3 \frac{x}{y}-\left(\frac{3}{2} f-\frac{x}{2 \bar{y}}\right) \ln \frac{x}{y}\right.\nonumber \\
&& \left.-(f-\bar{f}) \ln \frac{x}{y} \ln \left(1-\frac{x}{y}\right)+\left(f+\frac{x}{2 \bar{y}}\right) \ln ^2 \frac{x}{y}\right\}-\frac{x}{2 \bar{y}} \ln x(1+\ln x-2 \ln \bar{x})\nonumber \\
&& +\left\{\begin{array}{l}
x \rightarrow \bar{x} \\
y \rightarrow \bar{y}
\end{array}\right\}\,,\\
V_\beta(x,y)&= & \dot v(x,y)+\frac{5}{3}v(x,y)+v^a(x,y)\,,\\
V_G(x,y)&=& 2 v^a(x,y)+\frac{4}{3}v(x,y)+\theta(y-x)H(x,y)+\theta(x-y)\overline{H}(x,y)\,,
\end{eqnarray}
\end{subequations}
with $f\equiv f(x,y)$, $\bar f\equiv  f(\bar x,\bar y)$ and
\begin{subequations}
\begin{eqnarray}
\dot{v}(x, y)&=&\theta(y-x) f \ln \frac{x}{y}+\theta(x-y) \bar f\ln \frac{\bar x}{\bar y} \,,\\
v^a(x, y)&=&\theta(y-x) \frac{x}{y}+\theta(x-y) \frac{\bar{x}}{\bar{y}}\,,\\
 H(x, y)&=&2\left[\bar{f}\left(\mathrm{Li}_2(\bar{x})+\ln y \ln \bar{x}\right)-f \operatorname{Li}_2(\bar{y})\right]\,, \\
\overline{H}(x, y)&=&2\left[(f-\bar{f})\left(\operatorname{Li}_2\left(1-\frac{x}{y}\right)+\frac{1}{2} \ln ^2 y\right)+f\left(\operatorname{Li}_2(\bar{y})-\operatorname{Li}_2(x)-\ln y \ln x\right)\right]\,.\nonumber\\
\end{eqnarray}
\end{subequations}
In the above, $\bar x\equiv 1-x$ and $\bar y\equiv 1-y$. The $+$ distribution is defined through
\begin{subequations}
\begin{eqnarray}
	\int_0^1 dy \left[V(x,y)\right]_+ g(y)&\equiv& \int_0^1 dy V(x,y)\left[g(y)-g(x)\right]\,,\\\int_0^1 dx g(x) \left[V(x,y)\right]_+ &\equiv& \int_0^1 dx \left[g(x)-g(y)\right] V(x,y)\,,
\end{eqnarray}
\end{subequations}
for any test function $g(x)$.

\section{Some useful convolutions\label{appendix-convolutions}}

With the explicit expressions of the hard-kernel $T_H(x,m_Z,\mu)$, the LCDA $\hat\phi_H(x,m_Q,\mu)$ and the BL kernel $V^{(0,1)}_H(x,y)$, for $H=\eta_Q$, we have
\begin{subequations}
\begin{eqnarray}
 &&\int_0^1 d x T_P^{(0)}(x) \hat\phi_P^{(0)}(x)=1 \,,\\
&&\int_0^1 d x d y T_P^{(0)}(x) V^{(0)}(x, y) \hat\phi_P^{(0)}(y)=C_F(3-2 \ln 2) \,,\\
&&\int_0^1 d x d y d z T_P^{(0)}(x) V^{(0)}(x, y) V^{(0)}(y, z) \hat\phi_P^{(0)}(z)=C_F^2\left(2 \ln ^2 2-16 \ln 2-\frac{\pi^2}{3}+9\right)\,,
\end{eqnarray}
\end{subequations}
at the level of LLs,
\begin{subequations}
\begin{eqnarray}
 &&\int_0^1 d x d y T_P^{(1)}(x) \hat\phi_P^{(0)}(y)=C_F\left(\ln ^2 2-8 \Delta \ln 2+\ln 2-i \pi(3-2 \ln 2)-9\right)\,, \\
&& \int_0^1 d x d y T_P^{(0)}(x) \hat\phi_P^{(1)}(y)=C_F\left(8 \Delta \ln 2+2 \ln 2-\frac{\pi^2}{3}\right) \,,
   \end{eqnarray}
   \end{subequations}
at the level of NLO, and
\begin{subequations}
\begin{eqnarray}
&& \int_0^1 d x d y T_P^{(0)}(x) V^{(0)}(x, y) \hat\phi_P^{(1)}(y)\nonumber\\
&=&C_F^2\left[8\Delta  (\ln2-\ln^22)-3 \zeta (3)+\frac{2}{3} \pi ^2 \ln 2-2 \ln^22-\frac{4 \pi ^2}{3}-6
   \ln 2\right]\,,\\
 &&\int_0^1 d x d y T_P^{(1)}(x) V^{(0)}(x, y) \hat\phi_P^{(0)}(y)\nonumber\\
 &=&C_F^2\left[\frac{9 }{2}\zeta (3)-\frac{2}{3} \ln^32+\frac{\pi ^2 }{3} \ln 2+4 \ln
   ^22+23 \ln 2-27+8\Delta  \left( \ln^22- \ln 2\right)\right.\nonumber\\
   &&~~~~~~~~\left.-i \pi\left(2 \ln^2 2-\frac{\pi^2}{3}-16\ln 2+9\right)\right]\,, \\
   &&\int_0^1 d x d y T_P^{(0)}(x) V^{(1)}_P(x, y) \hat\phi_P^{(0)}(y)\nonumber\\
&=&C_F\Bigg[C_F\left(2 \zeta(3)+\frac{3}{2}-\frac{2}{3} \ln ^3 2+4 \ln ^2 2+4 \ln 2+\frac{\pi^2}{3}(2 \ln 2-1)\right)\nonumber\\
&& + C_A\left(-\frac{3}{2}\zeta(3)+\frac{4}{3} \ln 2+1\right)+\beta_0\left(\frac{1}{2}+\frac{\pi^2}{6}- \ln ^2 2-\frac{4}{3} \ln 2\right)-8 \beta_0 \Delta \ln 2\Bigg] \,,
\end{eqnarray}
\end{subequations}
at the level of NLLs, where $\Delta=0$ for the NDR scheme, and $\Delta=1$ for the HV scheme. One can see in the truncated expansion of the resumed SDC in Eq. (\ref{eq:P-NLL-expand}), the $\Delta$-dependence vanishes at each order of $\alpha_s$ as it should. The corresponding convolutions for $H=J/\Psi(\Upsilon)$ can be obtained similarly.


\end{document}